\documentclass[10pt,conference]{IEEEtran}
\usepackage{amsmath,amssymb,amsfonts}
\usepackage{algorithmic}
\usepackage{graphicx}
\usepackage{textcomp}
\usepackage{xcolor}
\usepackage[hyphens]{url}
\usepackage{fancyhdr}

\usepackage{mathptmx} 

\usepackage{enumitem}
\usepackage{multirow}
\usepackage[normalem]{ulem}
\usepackage[hyphens]{url}
\usepackage[sort,nocompress]{cite}

\usepackage[final]{microtype}

\usepackage[keeplastbox]{flushend}
\usepackage[bookmarks=true,breaklinks=true,letterpaper=true,colorlinks,citecolor=blue,linkcolor=blue,urlcolor=blue]{hyperref}
\usepackage{pifont}
\usepackage{stfloats} 
\usepackage{caption,subfig}

\newcommand{\cmark}{\ding{51}}%
\newcommand{\xmark}{\ding{55}}%

\usepackage[noabbrev,capitalise]{cleveref}
\usepackage{tikz}

\newcommand{\TRH}{$T_{RH}$}

\newcommand{\ignore}[1]{}

\def\BibTeX{{\rm B\kern-.05em{\sc i\kern-.025em b}\kern-.08em
    T\kern-.1667em\lower.7ex\hbox{E}\kern-.125emX}}
\definecolor{aliceblue}{rgb}{0.94, 0.97, 1.0}
\usepackage{xcolor}

\newcommand\arxivAdd[1]{#1}
\newcommand\arxivIgnore[1]{}

\definecolor{iris}{rgb}{0.35, 0.31, 0.81}

\Crefname{figure}{Fig.}{Figs.}
\crefname{figure}{Fig.}{Figs.}

\newif\ifvariance
\variancetrue
\usepackage{tcolorbox}
\tcbset{width=1\columnwidth,boxrule=1pt,colback=aliceblue,arc=1pt,left=2pt,right=2pt,boxsep=0.5pt}

\arxivIgnore{\title{{ {\bf \huge START: Scalable Tracking for Any Rowhammer Threshold} }}}
\arxivAdd{\title{{ {\bf \LARGE Scalable and Configurable Tracking for Any Rowhammer Threshold} }}}

\definecolor{myblue}{rgb}{0, 0, 1}
\newcommand\revhl[1]{%
  \bgroup  \hskip0pt\color{black!80!black}%
  #1
  \egroup
}

\author{
    \IEEEauthorblockN{Anish Saxena$^*$ and Moinuddin Qureshi}
      \IEEEauthorblockA{
        Georgia Institute of Technology \\
      }
}

\begin{document}
\maketitle
\arxivIgnore{\pagestyle{empty}}
\arxivAdd{\pagestyle{plain}}

\arxivAdd{\let\thefootnote\relax\footnote{$^*$The author can be reached at asaxena317@gatech.edu.}}
\begin{abstract}

The Rowhammer vulnerability continues to get worse, with the Rowhammer Threshold (\TRH) reducing from 139K activations to 4.8K activations over the last decade. Typical Rowhammer mitigations rely on tracking aggressor rows. The number of possible aggressors increases with lowering thresholds, making it difficult to reliably track such rows in a storage-efficient manner. At lower thresholds, academic trackers such as Graphene require prohibitive SRAM overheads (hundreds of KBs to MB). Recent in-DRAM trackers from industry, such as DSAC-TRR, perform approximate tracking, sacrificing guaranteed protection for reduced storage overheads, leaving DRAM vulnerable to Rowhammer attacks. Ideally, we seek a scalable tracker that tracks securely and precisely, and incurs negligible dedicated SRAM and performance overheads, while still being able to track arbitrarily low thresholds.

To that end, we propose \textit{START} - a \textit{Scalable Tracker for Any Rowhammer Threshold}. Rather than relying on dedicated SRAM structures, START dynamically repurposes a small fraction the Last-Level Cache (LLC) to store tracking metadata. START is based on the observation that while the memory contains millions of rows, typical workloads touch only a small subset of rows within a refresh period of 64ms, so allocating tracking entries on demand significantly reduces storage. If the application does not access many rows in memory, START does not reserve any LLC capacity. Otherwise, START dynamically uses 1-way, 2-way, or 8-way of the cache set based on demand. START consumes, on average, 9.4\% of the LLC capacity to store metadata, which is 5$\times$ lower compared to dedicating a counter in LLC for each row in memory. We also propose START-M, a memory-mapped START for large-memory systems. Our designs require only 4KB SRAM for newly added structures and perform within 1\% of idealized tracking even at \TRH of less than 100.

\end{abstract}

\section{Introduction}
\label{sec:intro}

DRAM scaling enables large-capacity memory that power the modern computing infrastructure.  With each scaling, DRAM cells become smaller and come closer to each other.  Unfortunately, such close packing leads to inter-cell interference. A prominent mode of this interference is {\em Rowhammer}~\cite{kim2014flipping, kim2014architectural}, wherein frequent activations to a DRAM row cause bit flips in nearby rows.  Rowhammer reamins a severe security threat~\cite{seaborn2015exploiting, frigo2020trrespass, gruss2018another, kwong2020rambleed, aweke2016anvil, cojocar2019exploiting, gruss2016rowhammer, van2016drammer, jang2017sgx,kwong2020rambleed}. For example, flipping bits in page tables leads to privilege escalation attacks.

\begin{figure*}
    \centering
     \includegraphics[width=2\columnwidth]{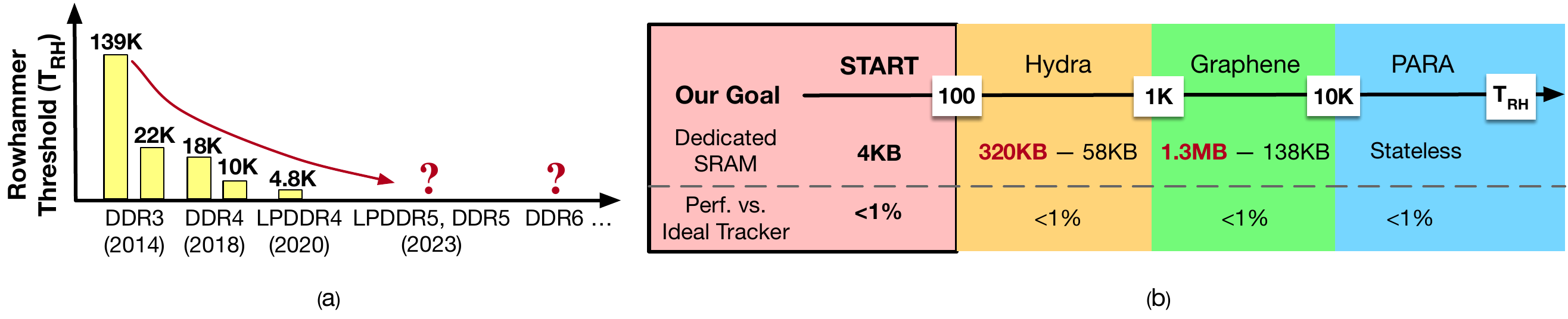}    
     \vspace{-0.1 in}
     \caption{(a) The trend of Rowhammer Threshold ($T_{RH}$) (b) The efficacy of various tracking mechanisms for our baseline 64GB memory system. Current solutions are not scalable to $T_{RH}$ of sub-100. Our proposed design, START,  efficiently tracks any $T_{RH}$.} 
    \label{fig:intro}
\end{figure*}

Alarmingly, Rowhammer keeps getting worse with each technology generation.  When the phenomenon was characterized in 2014, the {\em Rowhammer Threshold ($T_{RH}$)}, which denotes the activations required to an aggressor row within 64ms refresh period to induce a bit-flip in a nearby row , was 139K (for DDR3). As shown in Figure~\ref{fig:intro}(a),  $T_{RH}$ has steadily reduced, and the most recent study from 2020 reported $T_{RH}$ of only 4.8K (for LPDDR4). The $T_{RH}$ for current generation (DDR5) and future generation (DDR6) devices is expected to be much lower. Between 2014 and 2020, threshold reduced by 30X, and if the trend continues, we can expect sub-100 thresholds by the end of this decade.

As systems remain deployed for several years, effective Rowhamer solutions must handle not only current but also future thresholds. Our goal is to develop a practical and configurable Rowhammer defense that works for a range of Rowhammer thresholds. In line with prior works~\cite{qureshi2022hydra, saxena2022aqua}, we focus on low future thresholds of less than 500.

The typical solution for mitigating Rowhammer consists of (i) a {\em tracking mechanism} to identify an aggressor row (that is estimated to reach $T_{RH}$ activations), and (ii) a {\em mitigating action}, such as refreshing neighbouring victim rows. In this paper, our focus is the tracking mechanism. As threshold reduces, the number of rows that can become aggressors increases, therefore the required tracking resources  must increase in inverse proportion to the threshold (doubling when the threshold gets halved).  Tracking aggressor rows with low storage-overhead, low performance-overhead, and in a secure   manner has been a key subject of Rowhammer research,  as shown in Figure~\ref{fig:intro}(b).

At thresholds above 10K, tracking resources can be obviated by issuing mitigations probabilistically, as is done in PRA~\cite{kim2014architectural} and PARA~\cite{kim2014flipping}. However, probabilistic solutions start to incur considerable performance overheads due to unnecessary mitigations at lower thresholds. For thresholds at 10K and below, the performance overheads can be reduced by tracking and issuing mitigation when a row reaches $T_{RH}$ activations. An {\em Ideal Tracker}
provisions one SRAM counter for each memory row. However,
it incurs significant SRAM overheads. We note that
while memory system contains millions of rows, the fraction
of rows that are likely to reach TRH is still fairly small. Recent proposals dedicate SRAM tables to identify hot rows by
tracking only a small subset of rows, either at the memory controller (e.g. Graphene~\cite{park_graphene:_2020}) or inside
the DRAM-chip (e.g., Mithril~\cite{kim2021mithril}).
For example, the most storage-efficient tracker, Graphene,
requires 85KB for $T_{RH}$ of 8K for a 32GB DDR5 memory,
containing 4 million rows. Unfortunately, as the thresholds
reduce, the number of rows that can reach $T_{RH}$ increases, so
the number of rows to track also increases. For example, for
the same 32GB memory, at $T_{RH}$ of 1K and 256, Graphene
requires 680KB and 2.6MB, respectively.

Recent industrial solutions store the aggressor row counters in-DRAM (e.g. TRR~\cite{frigo2020trrespass},  DSAC-TRR~\cite{samsung_dsac}). 
Unfortunately, the tracker deployed in DDR4 does not track all aggressors and has been shown to be vulnerable~\cite{frigo2020trrespass}. In fact, recent white papers from JEDEC~\cite{JEDEC-RH1}\cite{JEDEC-RH2} clearly state that  “in-DRAM mitigations cannot eliminate all forms of Rowhammer attacks”, thus the systems remain vulnerable to Rowhammer attacks even in the presence of these in-DRAM TRR mitigations. Furthermore, recent research from industry on developing trackers for newer versions of in-DRAM TRR focuses mainly on doing so in an approximate manner in order to reduce storage overheads, while still suffering from significant escape probability, (for example, the recent DSAC-TRR~\cite{samsung_dsac} incurs 13.9\% probability of aggressor escaping detection between two mitigations), rendering such upcoming schemes insecure, and leaving future systems still vulnerable to Rowhammer attacks.

The SRAM overheads of tracking could be reduced by placing the tracking table (one counter for each row) within the DRAM and caching the entries on demand~\cite{kim2014architectural}. While thrashing of the small counter-cache can incur drastic performance loss, the recently proposed Hydra tracker~\cite{qureshi2022hydra} mitigates this by using an SRAM filter to track at a group-level until a subset of threshold is reached, followed by per-row tracking for rows above the subset threshold. At $T_{RH}$ of 256, Hydra incurs modest SRAM overhead of 93KB. However, at $T_{RH}$ of 64, Hydra incurs slowdown of more than 10\% with 93KB of SRAM per DDR5 channel. 
Ideally, we need a solution which {\bf (1)} precisely tracks activation counts at an arbitrarily low Rowhammer threshold, {\bf (2)} is configurable to any Rowhammer threshold without being restricted by the size of the dedicated structures decided at design time, {\bf (3)} incurs negligible SRAM overheads for newly added structures, and {\bf (4)} performs similar to idealized tracking. In this paper, we enable such a  solution.

We propose {\em Scalable Tracking for Any  Rowhammer Threshold (START)}, which precisely tracks activations of each row in memory, and is especially well suited to thresholds of 256 and below. We leverage the observation that applications that utilize the LLC well typically do not access a large number of rows within 64ms. Conversely, applications that access millions of rows within 64ms tend to have poor locality and are less sensitive to LLC capacity. Thus, the key insight of START is to obviate the dedicated SRAM storage of tracking by leveraging the last-level cache (LLC) to dynamically store the per-row counters. As typical workloads access only a small fraction of the memory rows within a period of 64ms, storage consumption is reduced significantly by tracking only the accessed rows. 

If the application does not access rows within 64ms, START does not reserve any LLC capacity. Otherwise, START dynamically allocates ways within a cache set as new rows are accessed. Each 64-byte line stores tracking metadata of 32 rows (including the row-tag). With 16MB of LLC capacity, reserving just 1-way across all sets can hold tracking entries of up-to 512K rows, which is sufficient in the common case (our system contains 64GB memory with 8 million rows). If a set requires more than 32 tracking entries, the allocation of that set is increased from 1-way to 2-way, and finally from 2-way to 8-way -- sufficient to track all 512 rows that map to the set with untagged counters. We require just 2-bits of state per set (4KB of SRAM, 0.02\% overhead) to track the per-set allocation. On average, START requires just 9.4\% of LLC capacity for counters, minimizing tracking-induced performance loss and performing within 1\% of an ideal per-row tracker.

For prior works, the structures for tracking the state of aggressor rows gets decided at the design time.  Thus, at design time, the designer must decide what would be the Rowhammer threshold during the system's lifetime, and this information may not be available.  It would be ideal if the solution can be configured (say at boot time) to the right Rowhammer threshold, without being constrained by the size of the dedicated structures.  While prior scheme do not allow such configurability, START enables it as the tracking state gets created dynamically based on need.  START uses a two-byte register, which is set at boot time, which allows START to track any threshold, while still performing within 1\% of an ideal tracker.

Finally, to support large-capacity memory systems, we propose {\em Memory-Mapped START (START-M)}, wherein the tracking data for all memory rows is stored in the DRAM and accessed only when the number of tracking entries exceeds the dedicated fraction of LLC capacity (8-ways per set). As START-M can store up-to 2.75 million tracking entries in the 8-ways of LLC, the likelihood of accessing memory for obtaining tracking entry becomes negligibly small. We evaluate START-M with 64GB of DRAM per core, and observe that START-M uses less than 12\% of LLC capacity for tracking, and performs within 1\% of an idealized tracker.

\vspace{0.05 in}

Overall, our paper makes the following contributions:
\begin{enumerate}
    \item To the best of our knowledge, we are the first to propose a configurable tracker which scales to sub-100 threshold.
   \item We propose START, which obviates the dedicated SRAM overheads of tracking by leveraging the LLC.
    \item We reduce the storage consumed for tracking by dynamically allocating per-set space based on demand.
    \item Our memory-mapped START design scales to large-memory systems and supports higher thresholds. 
\end{enumerate}
\newpage

\section{Background and Motivation}
\label{sec:background}

\subsection{DRAM Organization and Timing}\label{dram_org}

Modern DRAM-based memory is organized logically as channels, sub-channels, banks, and rows. In DDR5, each 64-bit channel consists of two independent sub-channels which are 32-bit wide with a burst length of 16 to supply a 64B line. Each sub-channel has  32 banks organized as a 2D array of rows and columns with 8KB row-size. The bank contains a {\em row buffer} that caches the most recently opened row.  To access data from DRAM, a row must be activated, which brings the data into the row buffer.
To access data in another row, the bank clears the row buffer using the {\em precharge} command, followed by activation of the given row. DRAM cells also require periodic refresh operations (at 64ms).

An important DRAM timing parameter is {\em $t_{RC}$ (Row Cycle Time)}, which determines the time between consecutive activations for a given bank. The $T_{RC}$ for DDR5 systems is approximately 45ns, which means a bank can encounter up-to 1.36 million activations $(ACT_{max})$ in the refresh window of 64ms, after discounting the time spent in refresh.

\subsection{Rowhammer and Security Threat}
\label{row_hammer}

\revhl{Rowhammer occurs when frequently activated rows cause bit-flips in nearby rows. {\em Rowhammer Threshold ($T_{RH}$)} denotes the number of activations required to any row, with any access pattern, to induce bit-flips in the nearby row.} When the Rowhammer phenomenon was first discovered in 2014, $T_{RH}$ was 139K, whereas it has reduced by 30X to 4.8K~\cite{kim2020revisiting} in 2020. $T_{RH}$ is likely to reduce even further for future DRAM technology. For example, if the trends hold, then a similar reduction of 30X would render a $T_{RH}$  of less than 100 by the end of the decade. Therefore, it is important that the solutions for Rowhammer are designed to tolerate not just the current $T_{RH}$ but also $T_{RH}$ for future nodes. 

Rowhammer poses a serious threat to system security and gives the attacker a powerful weapon to flip bits in Page-Tables for privilege escalation~\cite{zhang2020pthammer} or exploit the data-dependence of Rowhammer to read confidential data~\cite{kwong2020rambleed}.

\subsection{Threat Model}\label{threat_model}
We assume an unprivileged attacker that can run code natively on the system that is vulnerable to Rowhammer.
The attacker can run process(es) under {\em user} privilege and exploit Rowhammer to flip bits in the page-table or in another program's data to corrupt it\cite{zhang2020pthammer}. We assume the Rowhammer bit-flip occurs at the victim location when any row in memory incurs more activations than $T_{RH}$ within the refresh interval of 64ms. Thus, the attack is successful if no mitigation is issued when a row has encountered more than $T_{RH}$ activations.

\subsection{Scaling Challenges for SRAM Trackers}
\label{sec:pitfalls}

The typical method to mitigate Rowhammer is to track the activations and issue a {\em victim refresh} when a row reaches $T_{RH}$ activations. Prior studies have developed sophisticated algorithms to intelligently track aggressor rows by provisioning the tracking entries for only a small subset of memory rows.  

The minimum storage for tracking gets dictated by the number of rows that can encounter at-least $T_{RH}$ activations within the refresh period.  As $T_{RH}$ reduces, the number of rows that can reach the threshold increases, and hence the storage required for the tracking structures increases in direct proportion. In this paper, our goal is to develop a Rowhammer mitigation that works at thresholds lower than 500.  Table~\ref{table:motiv} shows the storage requirement of recent state-of-the-art trackers as $T_{RH}$ is reduced from 4K to 16, for a 64-GB memory.

\begin{table}[!htb]
  \centering
  \begin{footnotesize}
  \caption{SRAM/CAM storage required for 64 GB memory (two 32-GB DIMM, 128-Banks, 8KB-Row, 8M Rows).
  }
  \label{table:motiv}
\begin{tabular}{|c|c|c|c|c|}
\hline
$T_{RH}$     & \begin{tabular}[c]{@{}c@{}}Graphene\\ (CAM)\end{tabular} & \begin{tabular}[c]{@{}c@{}}DSAC-TRR\\ (CAM)\end{tabular} & \begin{tabular}[c]{@{}c@{}}Ideal Tracker\\ (SRAM)\end{tabular} & \textbf{\begin{tabular}[c]{@{}c@{}}Goal\\   \end{tabular}} \\ \hline \hline
16           & \textgreater 8 MB                                        & N/A                                                    & 4 MB                                                           & \multirow{5}{*}{\textbf{4 KB}}                                  \\ \cline{1-4}
64           & \textgreater 8 MB                                        & N/A                                                    & 6 MB                                                           &                                                                \\ \cline{1-4}
256   (target)       & 5.2 MB                                                   & N/A                                                    & 8 MB                                                           &                                                                \\ \cline{1-4}
1K           & 1.4 MB                                                    & 68 KB                                              & 10 MB                                                           &                                                                \\ \cline{1-4}
4K (current)  & 340KB                                                    & 16 KB                                               & 12 MB                                                           &                                                                \\ \hline \hline

Secure? & Yes & {\bf No} & Yes & Yes \\ \hline
\end{tabular}
\end{footnotesize}
\end{table}

\vspace{0.05 in}

\noindent{\bf Ideal Tracker (One-Counter-Per-Row)} represents the naive scheme that dedicates one SRAM counter for each row.  For a system with $R$ rows and threshold of $T_{RH}$, OCPR needs $R$ entries, each of $log2(T_{RH})$ bits. The storage requirement of the ideal tracker ranges from 12MB to 4MB, as the threshold is varied from 4K to 16. Note that the reduction in storage requirement at lower thresholds is due to smaller counters. Ideal trackers are traditionally considered impractical due to prohibitive storage requirements.

\vspace{0.05 in}
\noindent{\bf Graphene~\cite{park_graphene:_2020}} is the current state-of-the-art tracker. It uses the Misra-Gries algorithm to identify top-N frequently accessed rows, where $N$ is based on $T_{RH}$. While Graphene is quite effective at $T_{RH}$ of 4K (requiring only 340KB), the storage overhead of Graphene grows to more than 8MB at thresholds of sub-100. For example, at $T_{RH}$ of 64, provisioning 5-bit counter and 17-bit row-id for 40K potential aggressors takes up more SRAM (109KB per bank) than just storing 128K 5-bit untagged counters (80KB). This makes Graphene less appealing than the ideal tracker at sub-100 thresholds.

\vspace{0.05 in}

\noindent{\bf DSAC-TRR~\cite{samsung_dsac}} is a recent tracker proposed by Samsung. It combines a space-efficient tracking algorithm with stochastic insertions to minimize counters required to defend against \textit{known} adversarial access patterns that employ decoy rows. DSAC-TRR trades off security for area efficiency, with a 13.9\% probability of escape for aggressor between mitigations during an attack at $T_{RH}$ of 10K. Moreover, since the effective threshold of DSAC is $T_{RH}/2 - ACT_{tREFI}$, where $ACT_{tREFI}$ can be as high as 255, it does not scale to $T_{RH}$ below 500.

\begin{tcolorbox}

\noindent{\bf Key Takeaway:} At ultra-low thresholds, {\em intelligent} trackers like Graphene require significantly more SRAM storage than an idealized tracker (due to extra metadata and over-provisioning). To enable practical Rowhammer mitigations in this regime, we must make ideal tracking viable.  
\end{tcolorbox}

\subsection{Scaling Challenges for Hybrid Tracker}
\label{sec:hydrapitfalls}

The SRAM overheads of an ideal tracker can be reduced by placing the counter table within the DRAM and caching the entries on demand in a metadata cache, as proposed in {\em Counter-Based Row Activation (CRA)}~\cite{kim2014architectural}.  Unfortunately, even in presence of large metadata-caches (64KB-256KB), CRA experiences a significant number of extra accesses for fetching counter lines because of poor spatial locality, causing drastic slowdown (averaging 25\%~\cite{qureshi2022hydra}), limiting practical adoption.

A recent proposal, {\em Hydra}~\cite{qureshi2022hydra}, uses a hybrid design where an SRAM  filter reduces the DRAM accesses for per-row counters. Hydra contains an SRAM structure that performs aggregated tracking for a group of rows until a subset of the Rowhammer threshold is reached.  Per-row tracking is enabled only for rows for which the group-level threshold is breached.  Hydra at $T_{RH}=500$ was shown to have low overheads and slowdown. 

The SRAM overhead of Hydra depends on $T_{RH}$ and the number of channels. For our baseline system with two DDR5 DIMMs, Hydra incurs an SRAM overhead of 186KB for a threshold of 256.  However, as the threshold reduces, the number of entries in the Hydra SRAM structures must be increased in direct proportion (that is, 4X more entries if the threshold is reduced by 4X). So, at thresholds of 64 and 16, Hydra incurs an SRAM overhead of 544KB and 1.3MB, respectively. If the storage overheads are not increased, then Hydra incurs significant slowdowns at lower thresholds.

\begin{figure}[!htb]
   \vspace{-0.1 in}
    \centering
    \includegraphics[width=1\columnwidth]{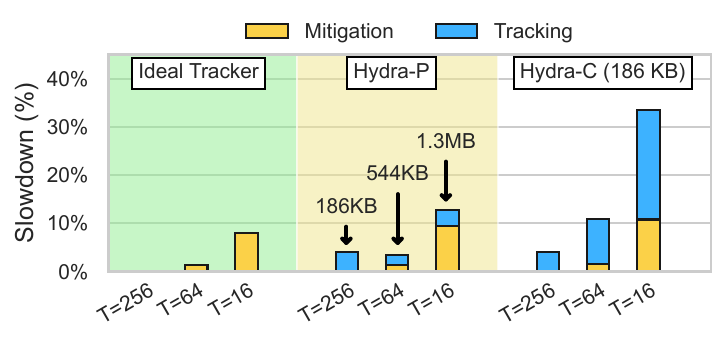}
   \vspace{-0.1 in}
    \caption{Slowdown of Ideal Tracker, Hydra-C (186KB), Hydra-P (proportional storage) for thresholds of 256 to 16. Hydra incurs a slowdown due to both mitigation and tracking. }
    \label{fig:hydra}
   \vspace{-0.1 in}
\end{figure}

Figure~\ref{fig:hydra} shows the slowdown of ideal tracker, {\em Hydra-P} (proportional storage), and {\em Hydra-C} (constant 186KB)  as $T_{RH}$  is varied from 256 to 16.  Ideal tracker incurs slowdown only due to mitigation, whereas Hydra suffers from both mitigation and metadata memory accesses. The overhead of ideal tracker due to mitigation alone is relatively small, 0.2\% at $T_{RH}=256$, 1.3\% at $T_{RH}=64$ and 8\% at $T_{RH}=16$, because modern memory devices have a large number of banks and concurrency, hiding the impact of victim refresh in a bank.  The overhead of Hydra-P is within 2\% of the ideal tracker. However, if we do not provide the proportional SRAM storage to Hydra, then the constant storage configuration (Hydra-C) incurs significant slowdowns, from 4.2\% at $T_{RH}=256$ to 34\% at $T_{RH}=16$. Thus, Hydra at sub-100 thresholds incurs either significant SRAM overhead or significant slowdown.

\subsection{Our Goal}

We observe that at ultra-low thresholds, existing proposals either require prohibitive SRAM overheads, or performance overhead, or both. Furthermore, for all prior proposals, the SRAM structures are provisioned to target a particular Rowhammer threshold, and this decision is taken at design time.  Therefore, the system becomes incapable of handling a memory module that is known to have a lower threshold. 
\begin{tcolorbox}
The goal of this paper is to develop a scalable tracking mechanism with the following attributes: (1) Precise row tracking at an arbitrarily low threshold (2) Configurable to a given threshold without being restricted by the size of the tracking structures (3) Incurs negligible SRAM overheads for newly added structures, and (4) Incurs negligible slowdown compared to an ideal tracker.
\end{tcolorbox}

\section{Evaluation Methodology}
\label{sec:methodology}

\subsection{Simulation Framework} 

We use ChampSim~\cite{gober2022championship}, a cycle-level multi-core simulator, interfaced with DRAMSim3~\cite{li:dramsim3}, a detailed memory system simulator. We modified DRAMSim3 to include the DDR5 configuration, wherein each DIMM supports two sub-channels that can be operated independently and provides a 64-byte line with a burst length of 16.  We use the DRAM-based power model provided by Micron~\cite{micron_ddr5}. Table~\ref{table:system_config} shows the configuration for our baseline system. 

\begin {table}[h!]
\begin{footnotesize}
\begin{center} 
\vspace{-0.05 in}
\caption{Baseline System Configuration}
\vspace{-0.05 in}

\begin{tabular}{|c|c|}
\hline
  Out-of-Order Cores           & 8 cores at 4GHz       \\
  ROB size           & 352       \\
  Fetch, Dispatch, Retire width & 6, 6, 5         \\ 
  L1-I/D and L2 (Private) & 32KB and 512KB, 8-way \\ \hline
  Last Level Cache (Shared)    & 16MB, 16-Way, 64B lines, SRRIP \\ \hline
  Memory size                  & 64GB -- DDR5 \\
  Memory bus speed             & 2.4 GHz (4800 MT/s) \\
  t$_{RCD}$-t$_{CL}$-t$_{RP}$-t$_{RC}$ & 16.6~-~16.6~-~16.6~-~48.6 ns\\
 Channels                      & 2 (one 32GB DIMM per channel) \\  
  Banks x Ranks x Sub-Channels & 32$\times$1$\times$2 \\
  Rows per bank                & 64K \\ 
  Size of row                  & 8KB  \\ 
  Sub-Channel width and BL     & 4B and 16   \\ \hline
\end{tabular}
\label{table:system_config}
\vspace{-0.1 in}
\end{center}
\end{footnotesize}
\end{table}

\revhl{We evaluate performance using 8 out-of-order cores with private L1 and L2 caches and shared L3 cache. The L3 is non-inclusive, with 128 MSHRs/core, 32 entry/core read and write queues, 4 read and write ports, 30-cycle hit-latency, no prefetcher, and SRRIP replacement policy. Our memory system contains two channels, each with a 32GB DDR5 DIMM (total of 64GB containing 8 million 8KB rows). }

For evaluating prior Rowhammer mitigation schemes, all SRAM structures associated with tracking are incorporated into the memory controller. For the mitigating action, without loss of generality, we assume  victim refresh of one neighboring row on each side using {\em Directed Refresh Management (DRFM) command}, where the memory controller supplies the aggressor row address to the memory, and the memory internally refreshes the victims rows. Unless specified otherwise, we assume a default Rowhammer threshold of 256.

\subsection{Workload Characterization}
\label{sec:wc_characterization}
We evaluate our design using the publicly available ChampSim traces, which includes 10 from SPEC2017~\cite{SPEC2017}, 13 from LIGRA~\cite{shun:ligra} (graph processing), and  5 from PARSEC~\cite{bienia2008parsec}. These traces have been collected after fast-forwarding the workload to a region-of-interest. We perform  a warm-up period of 50 million instructions for each workload. \revhl{Eight copies of the same workload runs on 8 cores and continue executing until all 8 cores complete 200 million instructions each. }

Table~\ref{table:wc} shows the key characteristics of our workloads, including the average per-core IPC, the  LLC-Misses Per 1000 Instructions (MPKI), the workload footprint (calculated as the unique 4KB pages touched by the workload), and Unique-Rows touched within a period of 64ms, on average. The last row of the table captures the geometric mean of the IPC and the arithmetic mean of other values across all 28 traces.

\begin{table}[!htb]
  \centering
  \begin{footnotesize}
      
  \caption{Workload Characteristics: IPC, MPKI, footprint, and Unique Rows Touched (average within 64ms).}
  \label{table:wc}
  \begin{tabular}{|c|c|c|c|c|}
    \hline
Workload    & IPC	&	MPKI	& Footprint &	Unique Rows	 \\ 
            & (per-core)	&	(LLC)	& (8-core)   &	Touched (64ms) \\ \hline \hline
fotonik3D   & 0.49  & 19.7  & 16.1 GB   & 2,126K   \\
mcf         & 1.1   & 14.4  & 4.7 GB    & 1,170K   \\
gcc         & 0.31  & 17.8  & 1.9 GB    & 184K    \\
omnetpp     & 0.53  & 10.9  & 1.6 GB    & 396K      \\
bwaves      & 0.67  & 14.4  & 1.2 GB    & 260K      \\
roms        & 0.89  & 6.2   & 511 MB    & 130K       \\
cactuBSSN   & 1.59  & 7.8   & 473 MB    & 121K      \\
wrf         & 0.83  & 11.7  & 277 MB    & 71K        \\
pop2        & 1.92  & 3.5   & 219 MB    & 56K       \\
xalancbmk   & 1.12  & 2.1   & 157 MB    & 40K        \\
\hline \hline
CF          & 1.08  & 9.6   & 2.7 GB    & 677K       \\
BC          & 0.47  & 31    & 2.2 GB    & 438K       \\
PR-Delta    & 0.41  & 24.6  & 2.1 GB    & 389K       \\
BFSCC       & 0.71  & 23.4  & 2 GB      & 471K       \\
BFS         & 0.67  & 19.5  & 1.9 GB    & 431K      \\
Radii       & 0.59  & 26.7  & 1.2 GB    & 220K       \\
Triangle    & 0.79  & 16.5  & 1 GB      & 258K      \\
Components  & 0.59  & 40.8  & 920 MB    & 162K       \\
Comp-SC     & 0.57  & 40    & 915 MB    & 162K      \\
PageRank    & 0.47  & 54.7  & 878 MB    & 151K      \\
BFS-BV      & 1.1   & 12.6  & 763 MB    & 194K       \\
BellmanFord & 1.09  & 8.4   & 751 MB    & 191K       \\
MIS         & 1.47  & 7.1   & 700 MB    & 178K       \\
\hline \hline
canneal     & 0.33  & 15    & 1.3 GB    & 301K       \\
fluida      & 0.88  & 6.7   & 789 MB    & 201K       \\
raytrace    & 1.11  & 5.7   & 453 MB    & 116K        \\
facesim     & 0.83  & 6.4   & 182 MB    & 46K       \\
streamc     & 1.04  & 13.6  & 69 MB     & 18K      \\
\hline \hline
Average & 0.76      & 16.8 & 1.7 GB    & 327K      \\ \hline
  \end{tabular}
 \vspace{0.05 in}
  \end{footnotesize}
\end{table}

\subsection{Figure of Merit}

The primary figure of merit for our evaluations is the normalized performance compared to an unprotected baseline. To estimate the system performance, we measure the average IPC across all the 8 cores. As all cores run the same workload, the IPC variation across the cores is small.  

We also consider secondary metrics such as (a) SRAM overhead for newly added structures, (b) loss in LLC capacity, (c) change in LLC misses, (d) impact on system power consumption of DRAM and the LLC, (e) sensitivity to LLC capacity, and (f) the blast radius of the mitigation.
\newpage
\section{Scalable Rowhammer Tracking}
\label{sec:design}

To enable practical Rowhammer mitigation at ultra-low thresholds, we propose {\em Scalable Tracking for Any Rowhammer Threshold (START)}. START performs precise tracking of row activations of memory rows without requiring any new SRAM structure for storing the tracking entries.  The key insight of START is to obviate the dedicated SRAM storage of tracking, by leveraging the last-level cache (LLC) to store the per-row counters. For our baseline system with 64 GB memory (8 million rows), even with a 1-byte counter per row, we would need 8MB of space to store the counters. 

A naive design of START, which we call {\em START-Static or START-S}, simply reserves 8 ways of the 16MB 16-way LLC to reserve the counters.  However, doing so reduces the LLC capacity considerably, causing a significant slowdown. Therefore, we develop a dynamic scheme, which we call {\em START-Dynamic or START-D}, which adaptively allocates tracking entries only for the rows that get accessed within 64ms. We observe that only a small fraction of memory rows (about 4\% on average) get touched within 64ms, so START-D consumes significantly less storage. In this section, we provide an overview of START, using START-S as a simplified example, then describe START-D, and provide results and analysis.

\subsection{START-S: The Naive Design}

\begin{figure}[!htb]
    \centering    
    \includegraphics[width=0.8\columnwidth]{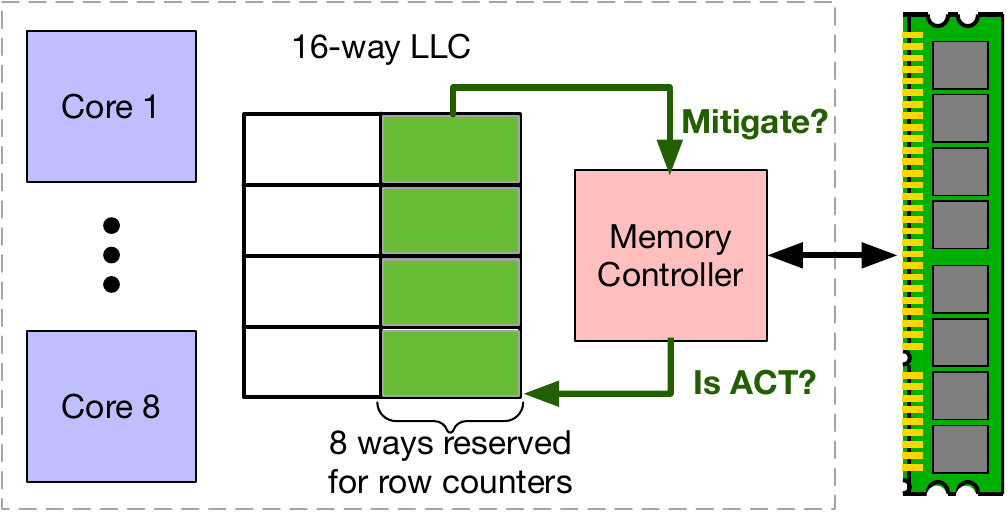}
    \caption{Overview of START using START-S.}
    \label{fig:starts}
   \vspace{-0.1 in}
\end{figure}

Figure~\ref{fig:starts} shows an overview of START-Static (START-S) design. Even though  START-S is inefficient, we use it to provide an overview of START owing to its simplicity. START-S reserves 8 ways of the 16-way LLC to store the counters for the 8 million rows. Let the ways reserved for storing the counters be ways 0 through 7. Then, on an LLC miss, these ways do not participate in the LLC replacement algorithm, so these lines cannot be removed from the cache.

With 1-byte counter per row, each cache line of 64 bytes stores the tracking entries for 64 rows. As each row has a dedicated tracking entry, these entries are untagged. To obtain a tracking entry for a given row, we hash the row address to the cache set, then use 3-bits of the row address to select one of the reserved ways, and then use 6  bits of the row address to identify the byte-in-line that stores the tracking information.

When a demand access probes the LLC and encounters an LLC miss, it gets routed to the memory controller to perform DRAM access.  If this access results in row activation, the memory controller provides the row address to the cache controller, so that the controller can obtain the tracking entry and increment the counter. If the counter reaches the threshold, the counter is reset and a signal for performing mitigation for the given row is provided to the memory controller.

START-S consumes half of the LLC for Rowhammer tracking, therefore, it incurs significant performance overheads (on average 7.4\% in our evaluations).  We observe that while the memory system contains 8 million rows, a workload would typically not touch all these rows within the refresh period of 64ms.  In fact, based on the workload characterization in Table~\ref{table:wc}, we observe that on average about 300K rows get touched within 64ms (4\% of the total memory rows) and only 3 out of the 30 workloads touch more than 500K rows. If we could provide the space only to the rows that get touched at least once during the 64 ms period, then we can greatly reduce the storage consumption of tracking.  Our dynamic design, START-D, achieves this goal.

\subsection{START-D: The Optimized Design}
\label{sec:start_overview}

\begin{figure}[!htb]
   \vspace{-0.1 in}
    \centering    
    \includegraphics[width=0.9\columnwidth]{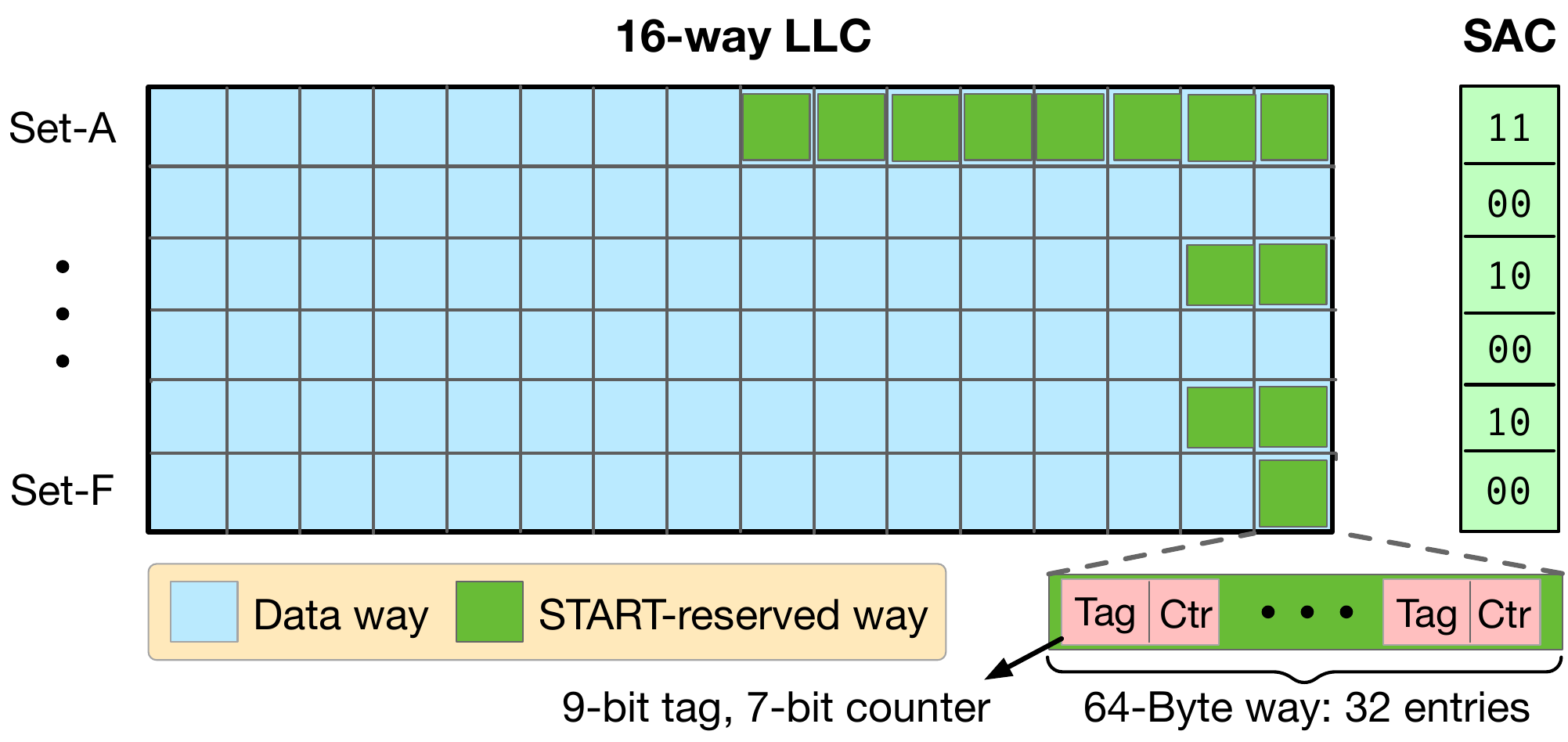}
   \vspace{-0.05 in}
    \caption{Dynamic allocation of START-D.  The number of ways reserved for each set varies based on demand.}
    \label{fig:startd}
   \vspace{-0.05 in}
\end{figure}

START-Dynamic (or START-D) varies the number of ways reserved for the tracking entries based on demand.  Figure~\ref{fig:startd} shows the overview of START-D. At the start of every 64ms period, START-D reserves no ways in the LLC and if the application does not access memory within this period (as the working set might already be cached), no LLC capacity is consumed. Otherwise, tracking entries get allocated on demand, and initially we use a tagged entry that identifies the row and the counter value. For our memory system with 8 million rows (23-bit row-id) and a cache with 16K sets (14-bit set index), the row-tag is 9-bits.  Without loss of generality, we use a 7-bit counter, to form a 2-bye tracking entry. Thus, a 64-byte line can store up to 32 tracking entries.

When a set in LLC receives the first request to update a row-counter mapped to it, the way-allocation is increased (from 0) to 1-way, thereby enabling storage of up-to 32 tagged counters. If all sets in LLC transition to this state, START-D can hold up-to 512K tracking entries. As only 3 workloads (fotonik3D, mcf, and CF) out of 30 touch more than 500K rows within a period of 64ms, this state is sufficient for most workloads in the common case, reducing the storage consumption of tracking by 8X (from 8-ways reserved to 1-way reserved).

If the LLC encounters a request for updating the counter for a given row, and all the tracking entries of the given set are in use, the allocation for that set is increased from 1-way to 2-way. The entries already stored in the first way are rehashed such that the even entries are retained in the first way and the odd tag entries are placed in the second way. The incoming entry is then allocated into one of the two ways depending on the row address (even tag or odd tag). Finally, in rare cases, if two ways are insufficient, then the allocation is increased to 8-way.  Note that 8-ways are sufficient to hold all tracking entries for the 512 rows that map to the given set, and all the tracking entries of the set are read and restored in an untagged format (each row has a designated byte for its tracking entry). Such reorganizations are also rare and not in the critical path.

\subsubsection{Newly Added Structure: SAC Table}

\begin{figure}[!htb]
   \vspace{-0.1 in}
    \centering    
    \includegraphics[width=0.8\columnwidth]{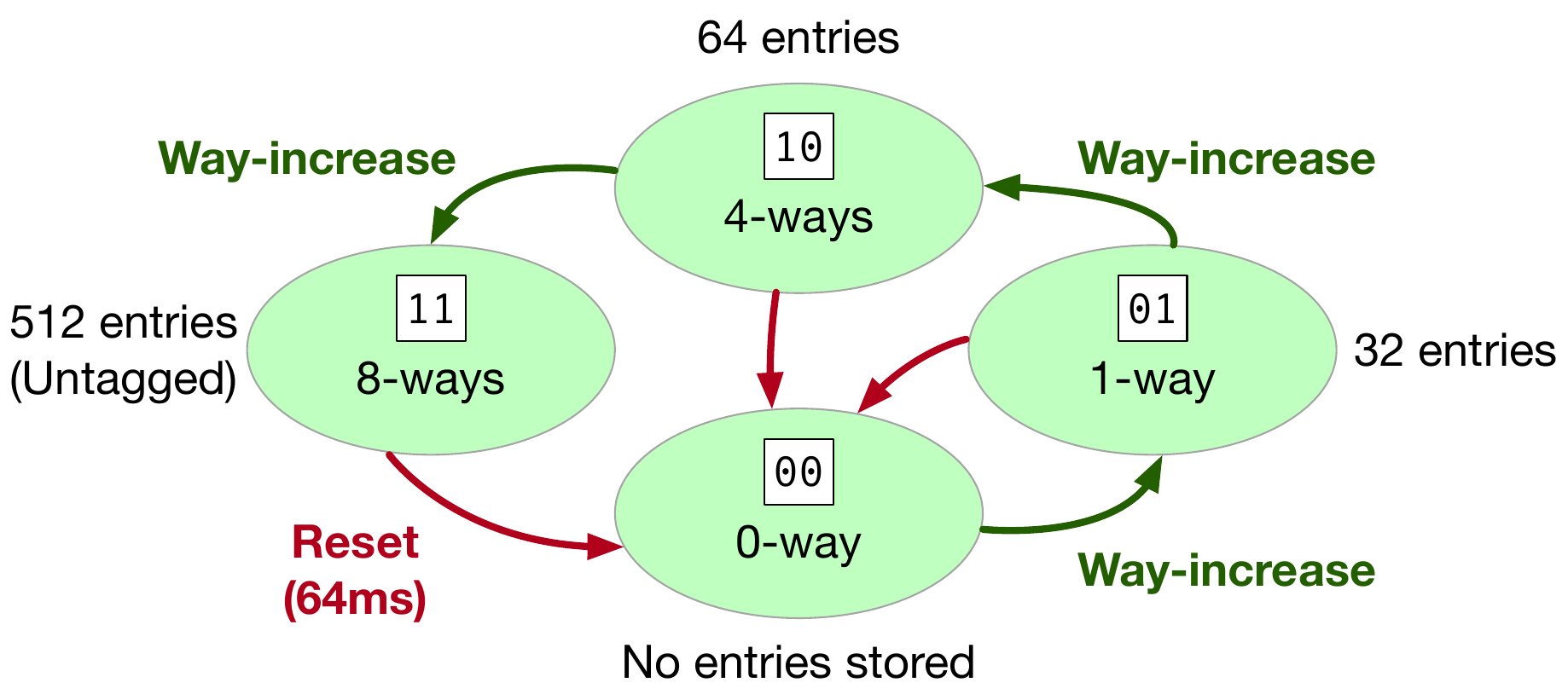}
   \vspace{-0.05 in}
    \caption{SAC transitions and the resulting allocation to a set.}
    \label{fig:sac}
\end{figure}

While START-D obviates dedicated SRAM for precise tracking information, it does require the state to indicate the number of ways that are reserved in each set to store tracking information.  Given that we have four possible allocations, we need two bits per set, which we call the  {\em Set Allocation Counter (SAC)}. If SAC is 00, the set has the default reservation of 0-way (no capacity reserved). If SAC is 01, the set has 1-way reserved (with 32 tagged entries). Similarly, if SAC is 10, the set has 2-ways of 32-entries each reserved (with each containing 32 tagged entries). Finally, if SAC is 11, the set has reserved 8 ways, and it would have 8 lines each storing 64 untagged entries, for a total of 512 entries. Figure~\ref{fig:sac} shows the transitions of SAC entries from 00 state to different states.  Every 64ms, the SAC entry of each set is reset to 00, so the allocation of a set remains valid only within the current refresh period. As each set requires a 2-bit SAC, and we have 16K sets, the table storing SAC entries (SAC Table) requires 4KB of SRAM. 

\subsubsection{Operations}

When the memory controller issues a row activation, it sends an update for that row to the LLC.  The LLC uses the top 14-bits of the 23-bit row-tag to identify the cache set that stores the given row's tracking entry. The cache controller checks the SAC entry for the set to find the number of ways reserved for the set.  If the SAC entry is 00, a way is allocated for tracking entries, and a tracking entry is allocated with the designated row-tag and counter value of 1. The SAC value is increased to 01. For subsequent row-updates to this set, the incoming row tag is compared with the row tag of all entries (for which the counter is nonzero). If the entry is found, the counter is updated. If the counter reaches the Rowhammer threshold, the counter is reset, and a signal is sent to the memory controller for issuing mitigation for the given row. If the entry is not found, then a new tracking entry is allocated, unless all 32 entries are allocated. In this case, SAC transition occurs, followed by tracking entry allocation in the appropriate way. To obtain the way-index, the-row-tag is hashed (1-bit, 3-bit for 2-way, 8-way respectively) and the process remains same as SAC value of 01.

\revhl{While START-D requires changes to the lookup and replacement policy of the cache, row-counter lookups are outside the critical path of demand accesses.}  On an LLC access or miss, the SAC of the given set is consulted, and depending on the SAC values, between 1 to 8 ways are removed from consideration of lookup or replacement. This also ensures that valid tracking entries do not get evicted from the LLC.

\begin{figure*}
   \vspace{-2mm}
    \centering
    \includegraphics[width=2\columnwidth]{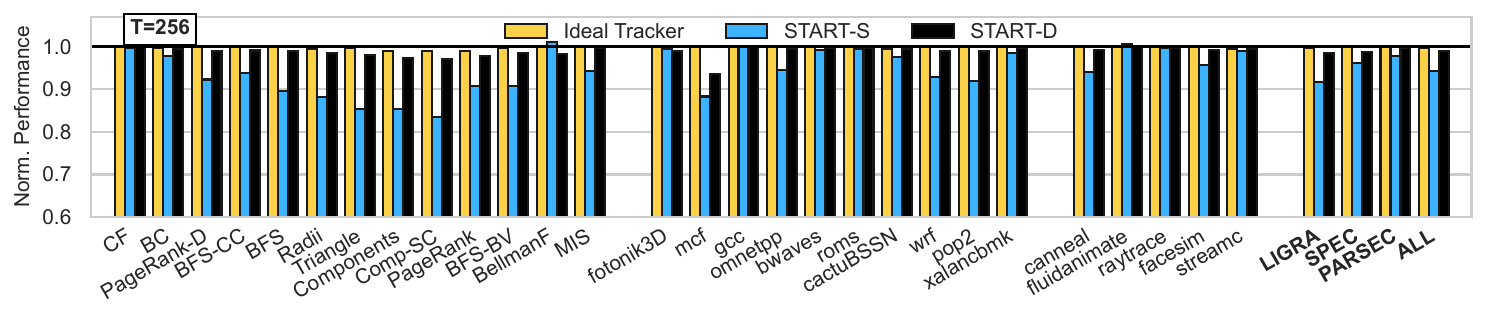}
   \vspace{-2mm}
     \includegraphics[width=2\columnwidth]{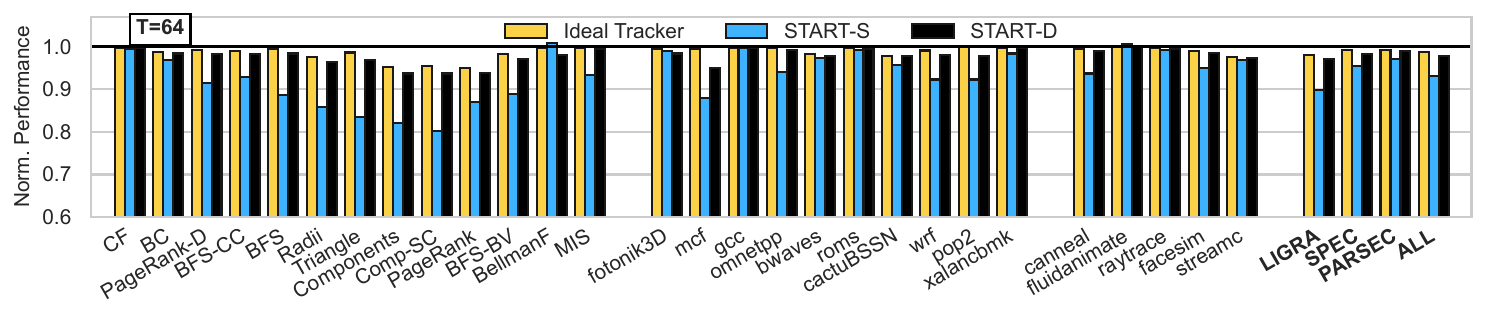}     
         \label{fig:start_t16}
            \vspace{-1mm}
    \caption{Performance of ideal tracker, START-Static, and START-Dynamic normalized to unprotected baseline. START-D performs within 1\% of an ideal tracker: slowdown of 1.1\% vs. 0.2\% at  $T_{RH}$ of 256 (top), and 2.2\% vs. 1.3\% at $T_{RH}$ of 64 (bottom).}
\vspace{-2mm}
    \label{fig:start_t64}

   \vspace{-4mm}
\end{figure*}

\subsubsection{Periodic Reset and Impact on Threshold}

We want to track the activation counts within 64ms, so, every 64ms, we reset the SAC table and the ways allocated in sets are released. This allows all ways to participate in cache replacement policy.

After SAC reset, the implicit row counts of all the rows are zero. As the reset of START-D may not be synchronized with refresh operations,  the attacker could potentially perform $(T-1)$ activations to the row before the reset and $(T-1)$ activations after reset and still not encounter any mitigation with a Rowhammer threshold set to $T$.  Thus, resetting causes the actual threshold tolerated by START to be $(2 \cdot T-1)$. Therefore, to tolerate a threshold of 256, we set the effective $T$ to be 128. This halving of effective threshold due to reset is a phenomenon that is common in prior trackers~\cite{qureshi2022hydra,park_graphene:_2020}.
\vspace{-0.05in}
\subsection{Impact on Performance}

\cref{fig:start_t64} shows the performance of START-S, START-D and Ideal Tracker normalized to the unprotected baseline. Ideal Tracker incurs only mitigation overheads and no tracking overhead. START-S suffers from considerable slowdown due to 50\% LLC capacity loss. In contrast, START-D closely follows the performance of ideal tracker for all workloads. The average slowdown of START-D is 1\% compared to 0.2\% for ideal tracker ($T_{RH}$ of 256, top). At $T_{RH}$ of 64 (bottom), START-D incurs 2.2\% average slowdown compared to 1.3\% with the ideal tracker (within 1\%).
START-D scales to arbitrarily low threshold, and performs within 1\% of an ideal tracker.

\subsection{Analysis on LLC Capacity Loss}

\begin{figure}[!htb]
   \vspace{-0.15 in}
    \centering
    \includegraphics[width=1\columnwidth]{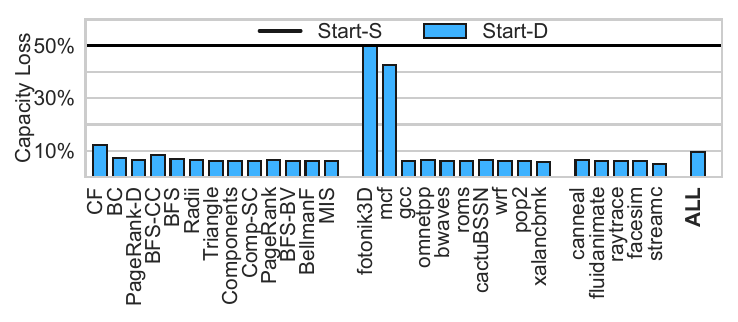}
   \vspace{-0.2 in}
    \caption{START-D requires 9.4\% of LLC capacity on average.
    }
    \label{fig:cap_loss}
   \vspace{-0.2 in}
\end{figure}

While START-S incurs constant 50\% capacity loss, the space consumed by START-D is proportional to the unique rows activated by the workload within 64ms (see \cref{table:wc}), as shown \cref{fig:cap_loss}. 
\revhl{On average, START-D only incurs 9.4\% capacity loss (5x lower than START-S), with 90\% of the workloads consuming less than 10\% of LLC capacity.}

\vspace{-3mm}
\subsection{Impact on Cache Misses}
\label{sec:cache_miss}

\begin{figure}[!htb]
    \centering
    \includegraphics[width=1\columnwidth]{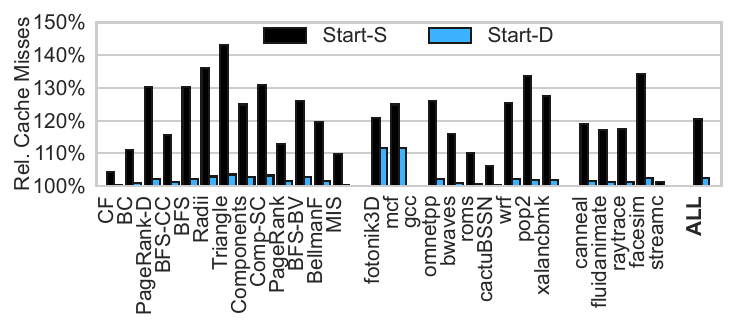}
   \vspace{-0.2 in}
    \caption{START-D increases Last Level Cache misses by just 2.3\% compared to the baseline, almost one-tenth of START-S.}
    \label{fig:cache_miss}
   \vspace{-0.15 in}
\end{figure}

\cref{fig:cache_miss} shows the increase in LLC misses due to START at $T_{RH}$ of 256. START-S significantly increases cache misses by 21\% on average. In contrast, START-D only incurs a negligible 2.3\% additional misses compared to the baseline.

\subsection{Sensitivity to Cache Size}
\label{sec:cache_sen}

\begin{figure}[!htb]
    \centering
    \includegraphics[width=0.9\columnwidth]{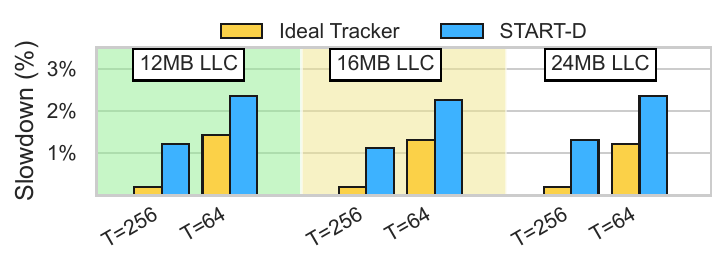}
    \caption{Impact on slowdown as LLC size is varied.  START-D performs similar to ideal tracker at different cache sizes.}
    \label{fig:cache_sens}
   \vspace{-0.1 in}
\end{figure}

\cref{fig:cache_sens} shows the performance of ideal and START-Dynamic trackers at different cache sizes compared to our default configuration (16MB, 16-way). In the non-default cache configurations, START-D dynamically reserves up-to 8-ways for 12MB 12-way LLC and up-to 4 ways for 24MB 12-way LLC. START-D incurs similar performance overheads, even at reduced cache sizes, because reservation of more than 1-way within a set remains exceedingly rare.

\subsection{Impact of Blast Radius}

\begin{figure}[!htb]
    \centering
    \includegraphics[width=0.9\columnwidth]{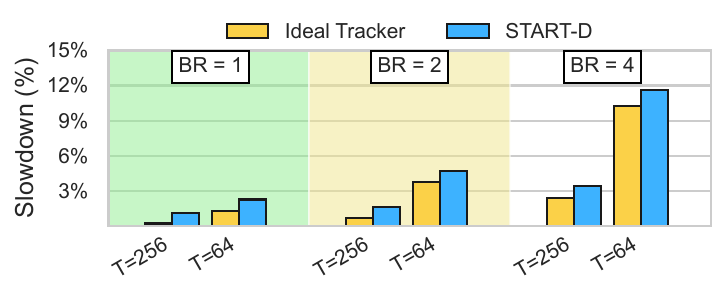}
   \vspace{-0.1 in}
    \caption{Slowdown with varying Blast Radius. START-D continues to perform within 1\% of ideal tracker.}
    \label{fig:blast_sens}
  \vspace{-0.1 in}
\end{figure}

Non-adjacent rows may also be impacted by activations to an aggressor row~\cite{half-double}. Recent proposals, therefore, increase the \textit{blast radius} of the mitigation by refreshing two or four adjacent rows on either side of the aggressor. We evaluate ideal and START-D with blast-radii from 1 to 4 in \cref{fig:blast_sens}. While the overheads of mitigation increase considerably with blast-radius, especially at $T_{RH}$ of 64, START-D maintains a slowdown similar to the ideal tracker with 11.2\% average slowdown compared to 10.2\% with the ideal tracker at BR=4.

\subsection{Storage and Power Overheads}
\label{sec:storage_power}

START-D requires 4KB SRAM for the SAC table (2 bits per set).
The size of SAC  depends only on the LLC capacity and not on the Rowhammer threshold. We also need two-bytes to store the Rowhammer threshold.

We use Micron's power calculator tool~\cite{micron:calc} to compute the DRAM power requirement. 
START-D increases DRAM power by 105mW at a negligible 0.3\% overhead. START incurs a LLC read and write for the row-counter on every DRAM activation. We compute SRAM power overheads using CACTI 7.0~\cite{CACTI} with 22nm technology. START-D incurs a dynamic cache power overhead of 93mW, an 11.5\% increase over baseline. However, taking the LLC leakage power into account, the overall cache power increases by only 0.9\%. 

\newpage
\subsection{Security Analysis}

For successful Rowhammer mitigation, START must ensure that it issues a mitigation before a row receives a threshold ($T_{RH}$) number of activations.  We define $T_{RH}$ as the minimum number of per-row activations to {\bf at-least} one row that are sufficient to cause a bit-flip via any attack pattern. To prove that START is secure, we make one assumption:

\begin{tcolorbox}
A successful row hammer attack requires activating {\bf at-least} one row more than $T_{RH}$ times within a refresh period.
\end{tcolorbox}

START is reset every 64ms. We call the period between consecutive reset as the {\em tracking window}. As  DRAM refresh is uncoordinated, a given DRAM row can experience two tracking windows within a single refresh period of 64ms.  So, START provides a stronger security guarantee, as follows:

\begin{tcolorbox}[boxrule=1pt,left=5pt,right=5pt,top=4pt,bottom=4pt]
\textbf{Theorem-1:} \textit{START issues mitigation for a row (a)~at $T_{RH}/2$ activations and (b)~at  each $T_{RH}/2$  activations since its past mitigation, in a tracking window.}
\end{tcolorbox}

\subsubsection{Proof of Security for Tracking by START}

Let $T_{true}$ be the exact or true count of a row's activations. We prove Theorem-1  analyzing two phases. Phase-1 is from reset to issuing the first mitigation. Phase-2 is between each consecutive mitigations.

In Phase-1, the activation counter entry associated with a given row is incremented whenever the row has an activation. So in Phase-1, the value of the counter is always equal to $ T_{true}$ of any row. Therefore, if the first mitigation for an aggressor row in a tracking-window is performed at $T_{RH}/2$, the activation count of the row ($T_{true}$) must reach $T_{RH}/2$. This proves part (a) of Theorem-1.  In Phase-2, the counter is reset to 0 upon a mitigation, and subsequently the tracking continues to be exact. The counter reaches $T_{RH}/2$ again only after performing $T_{RH}/2$ activations for the row after the mitigation. Therefore, the aggressor row is mitigated before receiving $T_{RH}/2$ activations (threshold is set to $T_{RH}/2$). This proves part (b) of Theorem-1. 

\subsubsection{Adaptive Attacks on START}

The attacker may try to dislodge the cache lines that store the tracking entry. However, this approach is not viable as the ways reserved for the tracking entries do not participate in the replacement algorithm. As LLC accesses due for tracking are outside the critical path of demand accesses, START does not introduce new timing side channels.

The mitigative action of performing refreshes of neighboring victims rows itself causes activations on victim rows. Recent Half-Double~\cite{half-double} attack exploits activations arising from refreshes of distance-1 neighbors to cause bit-flips in distance-2 neighbors. To be resilient to such attacks, START also includes any activation encountered due to victim refresh as part of the overall activation counts of the row.

We note that START is simply a tracking mechanism and can be use with any mitigating action.  We evaluate it with the commonly used mitigation of victim-refresh and a default blast radius of 1. If the DRAM modules have a higher blast radius, we assume that the mitigating action would be configured appropriately according to the blast radius.

\section{Memory-Mapped START (START-M)}
\label{sec:design_mm}

Thus far, we have considered baseline system with 8GB of memory-per-core (64GB for 8-cores) and 2MB LLC-per-core (total of 16MB) and $T_{RH}$ of 256 and below.
The tracking metadata for our system fits within a subset of the LLC (8MB) because each tracking entry is 2-bytes (9-bit row-tag and 7-bit counter) and the 512 untagged counters mapping to a set fit within 8-ways.
However, modern systems might have much larger memory capacity, or might work with current and old memory, which has a higher threshold than 256.
For example, an 8-core system with 512GB memory (64 GB-per-core) at threshold of 256 needs tracking metadata of 60MB, much larger than our 16MB LLC.
Similarly, if such a system operates at threshold of 4K (12-bit row-tag and 11-bit counter) would need more than 5X the LLC capacity of metadata.
START-D would be unable to handle such systems.
For such systems, we propose {\em Memory-Mapped START (START-M)}, which maintains a counter table in memory and uses tracking entries in the LLC to virtually eliminate all of the memory accesses for tracking$^1$\footnote{$^1$Multi-socket and disaggregated memory systems can be supported by provisioning tracking-entries in memory controller's parent socket.}. 
\subsection{Overview}

Consider a large-memory system with 512GB of memory and 8 cores (64GB of memory-per-core), operating at a threshold of 4K.
We maintain the other parameters similar to the previous baseline (16MB LLC, 2 DDR5 channels). 
As our baseline contains 64 million rows and 11-bit tracking entry for each row, it would require 82MB storage, well beyond LLC's capacity.

\begin{figure}[!htb]
    \centering    
    \includegraphics[width=0.9\columnwidth]{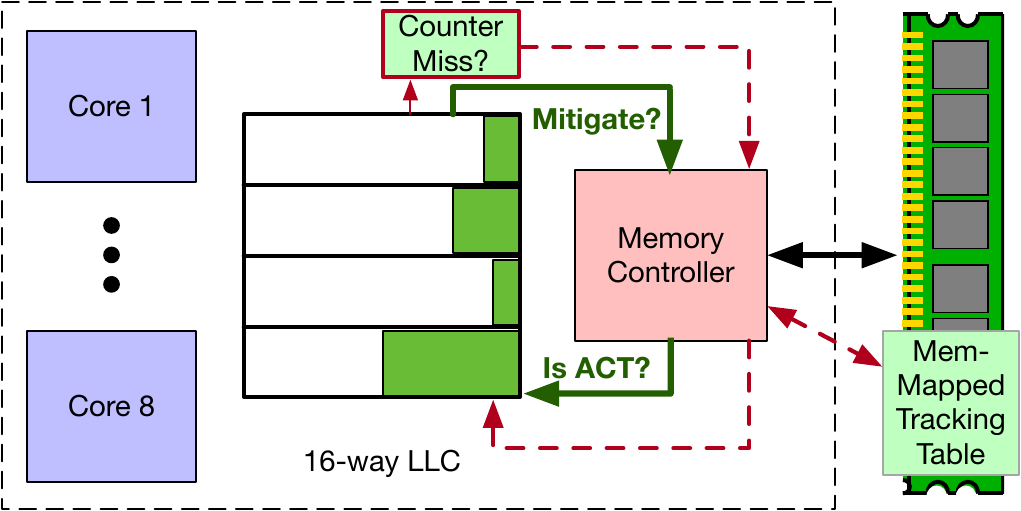}
    \caption{Overview of START-M. The dotted red arrows denote the rare case of metadata accesses to the DRAM.}
    \label{fig:startm}
\end{figure}

Figure~\ref{fig:startm} provides an overview of START-M. START-M reserves the required memory for untagged counters (82MB) in the addressable space of the main memory to store the {\em Memory-Mapped Tracking Table (MTT)}. But accessing the MTT to obtain the tracking metadata would require memory access and hence cause slowdowns. Rather than using a dedicated metadata  cache (as done in CRA~\cite{kim2014architectural}) or a filter (as done in Hydra~\cite{qureshi2022hydra}), START-M simply uses the LLC as the expandable area to store the tracking entries.  Similar to START-D, by default, START-M starts with no LLC capacity reserved (SAC of the set is set to 00). If a set requires entries, then the allocation is increased to 1-way, and then 2-ways, and finally 8-ways, on-demand, which is the maximum allocation allowed by our design. 

\subsection{Cache Changes: START-D to START-M}

The two key changes in START-M, compared to START-D are: (1) larger tracking entries, as memory capacity increases by 8x and counter size by 16x (12-bit row-tag and 11-bit counter), so each tracking entry in the LLC is 3 bytes (the changes from 1-way to 2-way now happens when there are 21 entries mapped to the set), and (2) always using tagged organization, as 8-ways (maximum allocation) are insufficient to hold the tracking entries for all the rows mapping to a set. Thus, even with 8-ways we use a tagged organization.

\begin{figure}[!htb]
    \centering    
    \includegraphics[width=0.9\columnwidth]{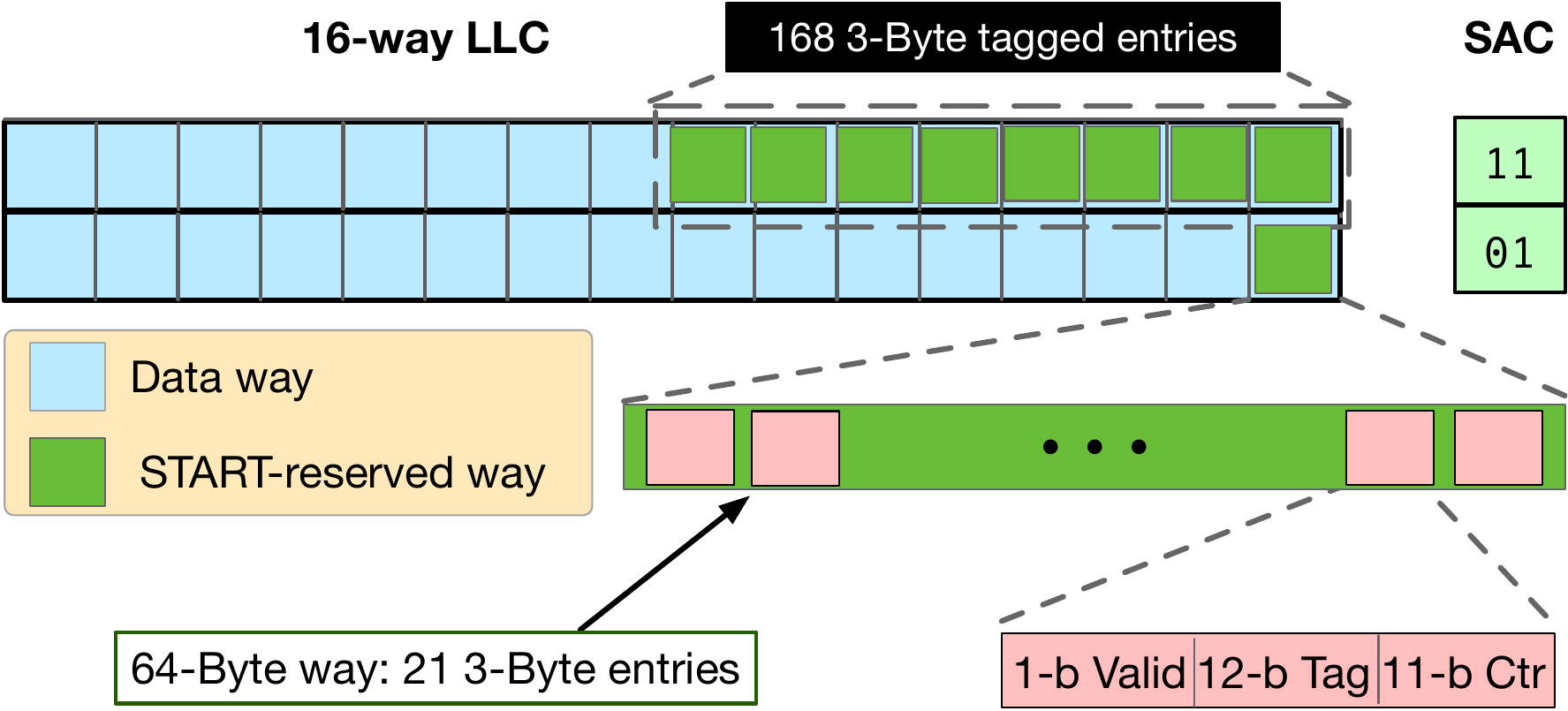}
   \vspace{-0.05 in}
    \caption{Set organization of START-M. Each tracking entry needs 3 bytes, and all allocations use tagged entries.}
    \label{fig:startm_org}
  \vspace{-0.05 in}
\end{figure}

If all sets are in state-1 (1-way reserved), START-M provides 344K tracking entries, which can increase to 2.75 million tracking entries at 8-ways. As workloads typically do not access these many unique rows within 64ms, virtually all of the memory accesses for obtaining tracking entries for START-M are cold misses (after the cache state is reset).  Next, we develop an optimization that avoids cold misses for the counters.

\subsection{Avoiding Cold Misses in START-M}
\label{sec:start_m_miss_optimization}

Every 64ms, the allocation of START-M reverts to 0-way reserved (SAC value of 00), similar to START-D. Thus, any unique row that gets accessed after the reset will not find the tracking entry in the LLC, and will access the memory. 

We leverage the observation that if there is space for the entry (e.g. allocation is less than 8-way or invalid tracking entries are present in the indexed way), then the given row is being accessed for the first time during the 64ms period. Otherwise, either the entry will be present, or the entry was evicted to accommodate another entry due to limited capacity. Therefore, we do not access the MTT on such {\em first-time} accesses and simply install the row in the LLC with a counter value of 1.
At reset, START-M also requires resetting the MTT in memory.  
We do this lazily by resetting all row-counters mapping to a set only when that set encounters its first-row eviction.
As each entry contains a valid-bit, the information to conduct this per-set reset is available without any extra overhead.
The episode of MTT accesses are also extremely rare, and we avoid the MTT reset overheads in the common case.

 With this optimization, START-M accesses the MTT only when there is no space for the tracking entry even with 8-ways, which cumulatively store approximately 2.75 million tracking entries in the LLC. As our workloads touch less than 2.2 million unique rows within 64ms, we observe negligible (less than 0.1\%) memory accesses for the MTT in our evaluations.  
\vspace{-0.5mm}

\begin{figure*}
    \centering
    \includegraphics[width=2\columnwidth]{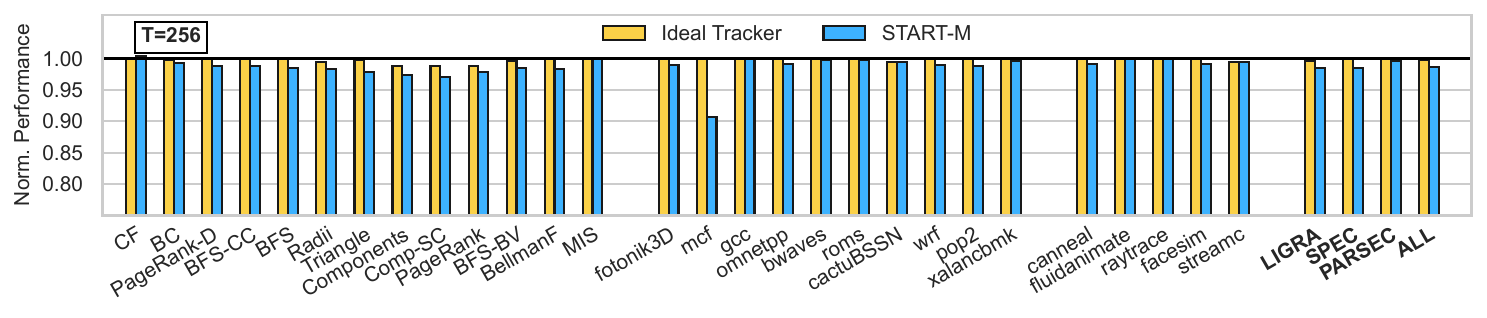}
    \label{fig:start_m_t64}
   \vspace{-2mm}

     \includegraphics[width=2\columnwidth]{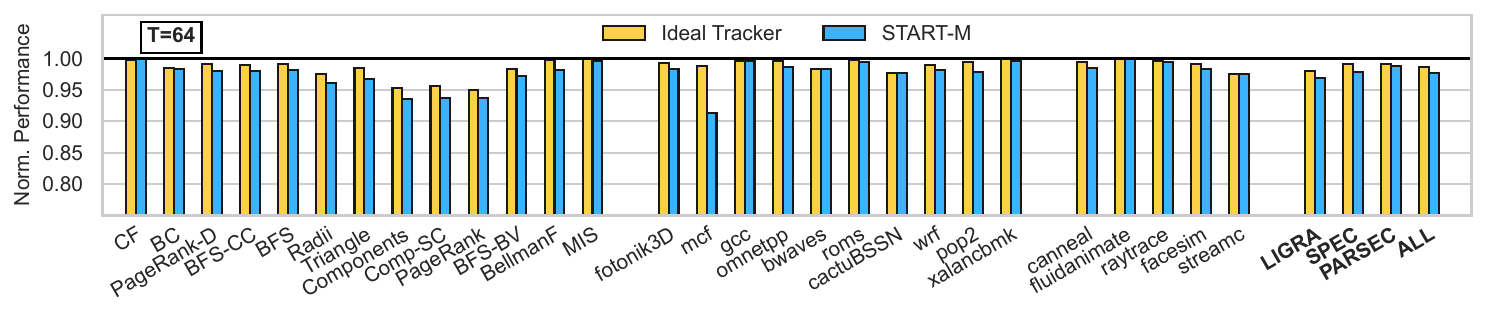}     
         \label{fig:start_m_t16}
            \vspace{-2mm}
    \caption{Performance of ideal tracker and START-M normalized to an unprotected baseline. START-M performs within 1\% of an ideal tracker: slowdown of 1.3\% vs. 0.2\% at $T_{RH}$ of 256 (top), and 2.3\% vs. 1.3\% at $T_{RH}$  of 64 (bottom).}

       \label{fig:startm_perf}

\vspace{-6mm}

\end{figure*}

\subsection{Impact on Performance}

Figure~\ref{fig:startm_perf} shows the performance of ideal tracker and START-M at thresholds of 256 and 64. At $T_{RH}$ of 256, START-M incurs slowdown of 1.3\%, similar to 0.2\% for ideal tracker. 
 At $T_{RH}$ of 64, START-M incurs an average slowdown of 2.3\% (within 1\% of the ideal tracker). START-M performs virtually identically to START-D while supporting a much larger memory capacity.

\subsection{Analysis of Cache Capacity Loss}

\begin{figure}[!htb]
   \vspace{-0.15 in}
    \centering
    \includegraphics[width=1\columnwidth]{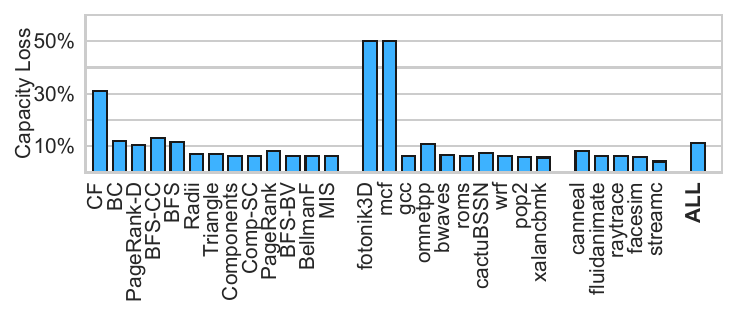}
   \vspace{-0.1 in}
    \caption{START-M requires just 11.4\% of the LLC capacity on average even with 1TB of memory provisioned per core.}
    \label{fig:start_m_cap_loss}
   \vspace{-0.15 in}
\end{figure}

\cref{fig:start_m_cap_loss} shows the loss in LLC capacity by START-M at $T_{RH}$ of 256. As START-M utilizes 3-Byte tagged counters and can store up-to 168 tagged counters within 8-ways, the cache capacity loss is 11.4\% compared to 9.4\% for START-D. All workloads, except \texttt{fotonik3D} and \texttt{mcf} (both access $>$1 million rows in 64ms), avoid memory accesses for tracking.

\subsection{Sensitivity to Rowhammer Threshold}

\begin{figure}[!htb]
    \centering
    \includegraphics[width=0.9\columnwidth]{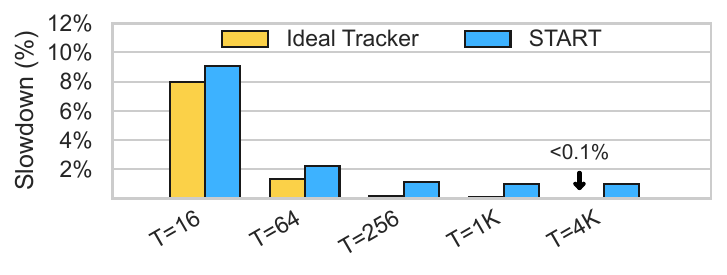}
  \vspace{-0.1 in}
    \caption{START scales from current threshold of 4K to extreme threshold of 16, remaining within 1\% of ideal tracker.}
    \label{fig:thres_sens}
   \vspace{-0.2 in}
\end{figure}

START's seamlessly scale to lower thresholds within a system's lifetime. ~\cref{fig:thres_sens} plots the overheads of START and ideal tracker as threshold is varied from 4K to 16. START-M is used for thresholds of 1K and 4K. START incurs 1\% overhead at $T_{RH}$ of 4K (ideal incurs negligible overhead). Even at the extremely low threshold of 16, START is within 1\% of ideal tracker with 9\% overhead compared to 8\% for ideal. 

\subsection{Security Considerations}

START-M accurately tracks the activation counts and issues mitigation when the count reaches $T_{RH}/2$, similar to START-D. START-M simply provides a larger backing store for the entries which does not affect the accuracy of tracking. As the accuracy of the per-row tracking is maintained, Theorem-1 is applicable to START-M. Nonetheless, placing metadata in memory opens up avenues for Rowhammer and denial-of-service attacks. An attacker can cause evictions of tracking entries from the LLC, making rows storing the tracking table (MTT) themselves aggressors and causing bit-flips in MTT. 
We maintain activation counts for rows storing the MTT (just like data rows) and issue mitigations when the activation count reaches $T_{RH}/2$. Thus, MTT-rows are protected against Rowhammer. 

While START-M virtually eliminates memory accesses for MTT for typical workloads, an adversarial workload can access several million rows randomly, thereby thrashing the tracking entries stored in the LLC. Applications would then cause 2X extra activations for each activation in the baseline (to obtain the tracking entry and to writeback the evicted tracking entry). The extra metadata activations are not in the critical path and do not cause latency overheads, but bandwidth overheads. For  bandwidth-limited systems, the adversary can cause performance degradation attacks even in the baseline by flooding the memory with requests. Memory system isolation solutions for such problems are also applicable to START-M.

\arxivAdd{\section{Discussion}

\subsection{Reduction of Rowhammer Threshold}

Over the past decade, the Rowhammer threshold has been characterized over thousands of DRAM devices~\cite{kim2020revisitingRH}, with a clear trend of significant threshold reduction over successive process nodes, as discussed in Section \ref{row_hammer}. Per that trend, it is likely that $T_{RH}$ of sub-100 will be reached in the next few years (or within the next decade). Sub-100 thresholds can be avoided if the DRAM organization changes fundamentally or DRAM vendors somehow mitigate Rowhammer. Unfortunately, even after a decade of efforts, neither option has materialized, as stated by JEDEC~\cite{JEDEC-RH1, JEDEC-RH2} and recent industry papers~\cite{samsung_dsac, isscc23}. For systems deployed today, chip designers must deploy defenses that work several years in the future on devices with currently unknown characteristics. To this end, START protects against arbitrary Rowhammer thresholds at low overheads, irrespective of whether they arrive in the next few years or the next decade. In both cases, our solution would be effective. 

\subsection{Pitfalls of Hybrid Tracking with Hydra}

Hydra tracks at a group-level until a group-threshold is reached, followed by row-level tracking by caching recently used row-counters in a dedicated cache. Unfortunately, the SRAM the filter and counter-cache must scale proportionately with increase in aggressors. Moreover, the dedicated SRAM structures must be provisioned at design-time. Hydra’s structure sizes depend on the range of thresholds (row-counter bits) and maximum memory supported (row-tag bits)~\cite{qureshi2022hydra}. For example, Hydra-544KB provisions 5-bit counters at threshold of 64, while 7-bit counters are needed support thresholds ranging from 64 to 256, requiring 700KB SRAM. Hydra is also not an exact tracker, as all row-group entries are initialized to the group-threshold when it is reached, even if many rows in the group encounter no activations, leading to spurious mitigations. 
Limited configurability, dedicated SRAM structures, and imprecise tracking limit Hydra's feasibility.
Table \ref{table:hydra_vs_start} compares Hydra with START-D at $T_{RH}$ of 64.

\begin{table}[!htb]
  \centering
 \vspace{-0.075 in}
  \begin{small}
  \caption{\revhl{Comparison of Hydra with START at $T_{RH}$ of 64.}}
  \label{table:hydra_vs_start}
  \setlength{\tabcolsep}{3pt}
  \renewcommand{\arraystretch}{1.15}
  \begin{tabular}{|c||c|c|}
    \hline
    \textbf{Attribute} & \textbf{Hydra-544KB} & \textbf{START-D}\\ \hline \hline
    Dedicated SRAM & 544KB & 4KB \\ \hline
    Memory-mapped Storage & 5MB & 0 \\ \hline
    Performance Overhead  & 3.2\% & 1.9\% \\ \hline
    SRAM Provisioned Dynamically  & {\xmark} & {\cmark} \\ \hline
    Scales to Arbitrary $T_{RH}$ & {\xmark} & {\cmark} \\ \hline
    Precise Tracking  & {\xmark} & {\cmark} \\ \hline
  \end{tabular}
  \end{small}
  \vspace{-0.15 in}
\end{table}

\subsection{A Case for Configurability via START}

Hydra incurs low performance overhead only if hundreds of KBs of SRAM is provisioned at design time, requiring additional chip area, power, and higher cost. 
As systems remain deployed for several years, designers must provision worst-case SRAM today for thresholds of the future. 
The dedicated storage can be rendered wasteful if lower thresholds are not reached within the system lifetime. Whereas, if ultra-low thresholds arrive in the absence of adequate dedicated storage, the system would experience a significant slowdown.

Our design solves the dichotomy with negligible dedicated SRAM overhead (4KB), while integrating with existing cache hierarchy, at negligible performance loss. Unlike Hydra, START can be configured for different use-cases \textit{at deployment}, for example, precise tracking within the LLC without a memory-mapped table with START-D or large-memory systems with higher thresholds up-to 4K with START-M (Section-V). 

In Appendices A to C, we extend evaluations to include multi-programmed and multi-threaded workloads for START, Hydra, and Ideal tracker, and present a new START policy that limits the LLC consumption to at-most 1 way.
}

\section{Related Works}
\label{sec:related}

\subsection{Tracking Row-Activation Counts}\label{sec:sram_mitigation}

The typical hardware-based solutions for mitigating Rowhammer rely on tracking the row activation counts and identifying aggressor rows.   This can be done either probabilistically, or with SRAM-based tracking or DRAM-based tracking.

\vspace{0.05 in}

\noindent{\bf Probabilistic trackers}, such as PARA~\cite{kim2014flipping} and PRA~\cite{kim2014architectural}, require significant mitigation overheads (almost 50X more compared to a precise tracker~\cite{samplingPRA}), which makes them unappealing at lower thresholds (less than 10K). Other schemes, such as MRLOC~\cite{MRLOC} and ProHIT~\cite{PROHIT} also use probabilistic decisions; however, they are not deemed secure.

\vspace{0.05 in}

\noindent{\bf SRAM-based trackers} use intelligent algorithms to track only a subset of rows in memory to identify the aggressor rows. Several such solutions exist, including Graphene~\cite{park_graphene:_2020}, CBT~\cite{CBT}, TWiCE~\cite{lee2019twice}, Mithril~\cite{kim2021mithril}, however all of them incur unacceptable SRAM/CAM overheads at ultra-low thresholds, sometimes even more (due to tag information) than an ideal tracker that dedicates one counter per row. D-CBF~\cite{yauglikcci2021blockhammer} use Bloom Filter for identification. However, to limit the false positives, such filters must be heavily over-provisioned, so they inucr significant SRAM overheads, and can only be used with rate control (and not victim refresh).

\vspace{0.05 in}

\noindent{\bf DRAM-based trackers} can provision one counter per row in a table that is resident in the DRAM.  However, accessing these counters can still incur significant overheads.  For example, Counter-Based Row Activation (CRA)~\cite{kim2014architectural} uses dedicated metadata caches to store these counters, however even with 64KB-256KB dedicated metadata caches it incurs a significant slowdown.  Hydra~\cite{qureshi2022hydra} is a recent proposal that uses SRAM filters (to do group tracking) to reduce memory accesses for tracking. While Hydra is effective at the thresholds of a few hundred, we show that it incurs significant SRAM overheads or slowdown at thresholds below 100.

\vspace{0.05 in}
\noindent{\bf Industrial trackers} such as {\em Target Row Refresh (TRR)} were developed by DRAM vendors to mitigate Rowhammer transparently. Recent attacks, such as TRRespass~\cite{frigo2020trrespass} and Blacksmith~\cite{jattke2021blacksmith}, exploit the fact that TRR keeps track of only a small number of aggressor rows and breaks TRR by issuing many requests.
Samsung's DSAC~\cite{samsung_dsac} and SK Hynix's PAT~\cite{isscc23} improve TRR for DDR5, but due to severe area limitation in DRAM, still allow aggressors to escape detection. DSAC has an escape probability of 13.9\% between two mitigations and PAT fails 6.9\% of the time (compared to DDR4-TRR). While area-efficient, these solutions are \textit{not} secure due to limited tracking capability.  
Thus, it is important to size the tracking structures considering the worst-case access pattern. START provides low-cost tracking for all the rows in memory.

\subsection{New Mitigating Actions for Rowhammer}

Our paper focuses on tracking activations; an orthogonal problem is mitigative actions.  We evaluate the mitigative action of {\em victim refresh}.  Recently, alternative mitigative actions have emerged, such as row migration (Randomized Row-Swap~\cite{saileshwar2022RRS}, AQUA~\cite{saxena2022aqua}, Scalable Row-Swap (SRS)~\cite{srs}, SHADOW~\cite{wi2023shadow}) and rate control (Blockhammer~\cite{yauglikcci2021blockhammer}). Of these, SRS caches heavily swapped rows (containing data), while we use LLC to store metadata (row-counters). Moreover, these solutions still need a tracker and can use START as a practical and scalable tracker, as START is compatible with any mitigative action.

\subsection{Modifying DRAM to Reduce Rowhammer}

Several recent  proposals modify the DRAM substrate to mitigate or reduce Rowhammer. For example, REGA ~\cite{rega} changes the DRAM substrate to generate extra refresh operations when a row is activated. SHADOW~\cite{wi2023shadow} modifies the DRAM microarchitecture to have an extra row with each sub-array and performs row migration to reduce Rowhammer.  {\em Panopticon}~\cite{bennett2021panopticon} proposes to redesign DRAM sub-array to store the counter alongside the DRAM row and increments this counter on each activation.  Our goal is to mitigate Rowhammer without needing to redesign DRAM arrays.  HiRA~\cite{hira}, can hide the refresh operations latency by refreshing a row concurrently with another access or refresh to the given bank. While HiRA may help with the mitigative action (such as victim refresh), it still needs a mechanism to identify aggressor rows.

\ignore{
\subsection{Modifying ECC to Tolerate Rowhammer}

Even memories that are equipped with Error-Correcting Codes (ECC) are vulnerable to Rowhammer attacks~\cite{cojocar2019exploiting}.  Recent proposals, such as SafeGuard~\cite{ali2022safeguard} and CSI:Rowhammer ~\cite{csi}, reorganize the ECC space to store not only the  {\em correction} metadata but also a {\em Message Authentication Code (MAC)}, which can be used to identify bit-flips due to Rowhammer. However, Rowhammer can still cause a large number of bit-flips, beyond the capability of the correction code, leading to data loss. Therefore, Rowhammer mitigation is still needed even in the presence of these solutions. 
}

\subsection{Virtualizing Predictors and Metadata}

START enables low-cost Rowhammer tracking by repurposing some of the LLC space for storing the tracking metadata. The paradigm of virtualizing a hardware structure by placing it in the cache space is quite powerful~\cite{lehman2018maps} and has been used in  prior academic and industrial proposals.  For example, such techniques have previously been applied to virtualize the prefetcher state~\cite{predvir} and the Branch Target Buffer (BTB)~\cite{phantomBTB}. AMD Magny-Cours processor uses part of the L3-Cache to store a probe filter~\cite{magnycours}.  However, our proposal (START-D) not only uses the concept of virtualization, but also dynamically allocates the storage required, to reduce the space consumption by almost 5X compared to a design that stores the full tracking table within the LLC (START-S).

\section{Conclusion}
\label{sec:conclusion}

As Rowhammer thresholds continue to reduce with each technology generation, we seek solutions that are effective for a range of current and future thresholds.  Tracking activation counts is a critical component of mitigating Rowhammer.  At ultra-low thresholds (sub-100) all prior tracking techniques incur either significant SRAM overheads, or performance overheads, or both.  In this paper, we propose {\em Scalable Tracking for Any  Rowhammer Threshold (START)}, which enables practical and precise tracking of row activations. START obviates dedicated SRAM overheads by placing the tracking metadata within the LLC, and uses a dynamic mechanism that allocates space on demand (thereby reducing the space needed by 5X). START requires only 4KB SRAM for newly added structures and performs within 1\% of an idealized tracker, even for thresholds of less than 100.

\begin{figure*}
   \vspace{-2mm}
    \centering
    \includegraphics[width=2\columnwidth]{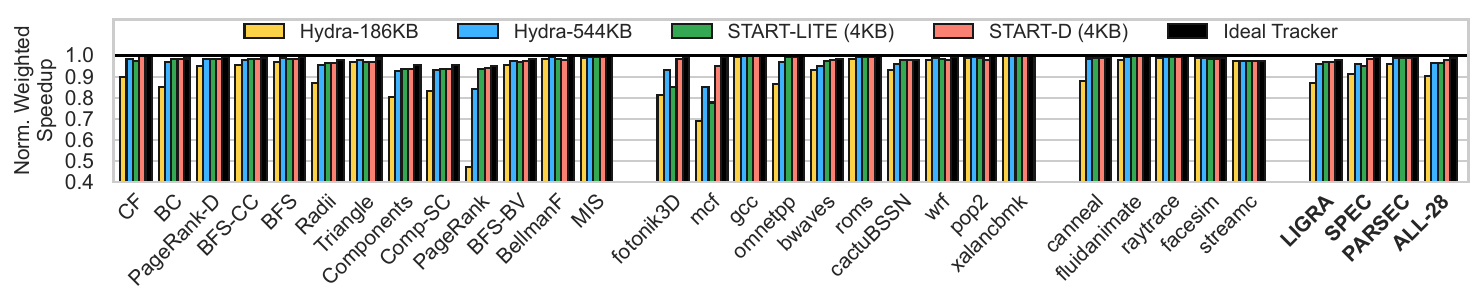}
   \vspace{-2mm}
     \includegraphics[width=2\columnwidth]{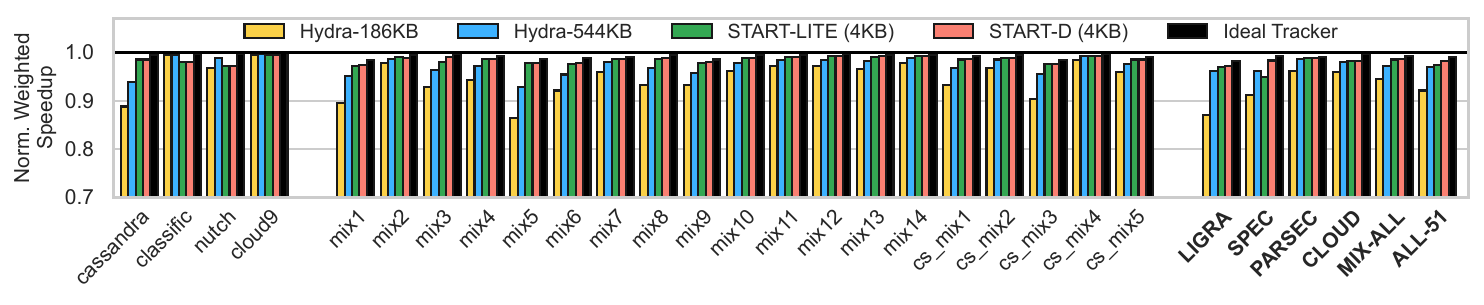}     
            \vspace{-1mm}
    \caption{\revhl{Performance of START, Hydra, and Ideal trackers normalized to unprotected baseline at $T_{RH}$ of 64 for single (top) and mix (bottom) 8-core workload configurations. On average, START-D incurs average slowdown of 1.9\% compared to 1\% for ideal tracker, while START-LITE incurs 2.7\% slowdown, similar to Hydra-544KB's 3.2\%, while requiring 136x less SRAM.}}
    \label{fig:start_lite_t64}

\end{figure*}
\section*{Appendices}

\section*{Appendix-A: Additional Workloads}
\label{app:more_eval}

In addition to 28 SPEC, LIGRA, and PARSEC workloads (details in Section \ref{sec:wc_characterization}), we evaluate multi-threaded, multi-programmed, and cache-sensitive workloads:

\vspace{0.05 in}
\noindent{\bf Multi-programmed Workloads:}  We generate 14 workload mixes by randomly selecting sets of 8 workloads from 28 SPEC, LIGRA, and PARSEC traces to run on 8 cores. We label them as \textit{mix1} to \textit{mix14}.

\vspace{0.05 in}

\noindent{\bf Multi-threaded CloudSuite Workloads:} We evaluate 4 CloudSuite workloads~\cite{cloudsuite} using two copies of the workload (4 unique traces per workload) running on 8 cores. Table \ref{table:cs_wc} shows characterisitics of the workloads, which are cache-sensitive (average LLC MPKI of 3.7 compared 16.8 for workloads in Table \ref{table:wc}). We also generate 5 mixes of CloudSuite, SPEC, LIGRA, and PARSEC workloads by randomly selecting 8 workloads from the 44 traces (labeled as \textit{cs\_mix1} to \textit{cs\_mix5}).

\begin{table}[!htb]
  \centering
  \begin{footnotesize}
 \vspace{-0.05 in}

  \caption{\revhl{CloudSuite Workload Characteristics.}}
  \label{table:cs_wc}
  \begin{tabular}{|c|c|c|c|c|}
    \hline
Multi-Threaded    & Weighted	&	MPKI	& Footprint &	Unique Rows	 \\ 
Workload            & Speedup	&	(LLC)	& (8-core)   &	Touched (64ms) \\ \hline \hline
cassandra   & 4.21  & 6.9  & 1.4 GB   & 365K   \\
classification         & 4.4   & 2.8  & 373 MB    & 95K   \\
nutch         & 5.42  & 3.1  & 203 MB    & 52K    \\
cloud9     & 4.65  & 2.2  & 110 MB    & 28K      \\
\hline \hline
Average & 4.65      & 3.7 & 528 MB    & 135K      \\ \hline
  \end{tabular}
  \end{footnotesize}
\end{table}

\section*{Appendix-B: START-LITE: Limiting LLC Usage}
\label{app:start_lite}

START-D can dynamically allocate up-to 8-ways (50\% of LLC capacity) for tracking entries. Each LLC access consults the Set Allocation Counter (SAC), so START's maximum allocation can be lowered by limiting the maximum SAC value. This is especially useful if START is co-running with other optimizations that require LLC resources, like way-partitioning or Data Direct I/O \cite{alian2020data}. As all tracking entries cannot fit in the LLC (in the worst-case), memory-mapped START can be used to store counters for each row in the memory and access them on-demand, while avoiding cold counter misses (Section \ref{sec:design_mm}).

We evaluate such a design, termed START-LITE, where maximum SAC value is \texttt{01} (1-way reserved), requiring just 6.25\% of LLC capacity in the worst case. START-LITE accesses the memory for metadata only when there is no space in the allocated way (32 tracking entries). Despite 8x lower LLC allocation than START-D in the worst case, the overhead is low because the evaluated workloads activate about 330K rows within 64ms on average (cf. Table \ref{table:wc}) and 1-way allocation in the LLC (Set Allocation Counter value of state-1) accommodates up-to 512K row-counters, making metadata memory accesses infrequent, as we show next.

\section*{Appendix-C: Slowdown of START-D and START-LITE}

Figure \ref{fig:start_lite_t64} shows the weighted speedup of START-D, START-LITE, Hydra and Ideal tracker normalized to an unprotected baseline at $T_{RH}$ of 64. Hydra with 186KB of SRAM incurs a significant slowdown (8.6\%) that reduces to 3.2\% by provisioning 3x more dedicated SRAM (Hydra-544KB). 

START-LITE requires only 4KB of dedicated SRAM and incurs only a 2.7\% slowdown. START-D further reduces the slowdown to 1.9\% (within 1\% of ideal) and does not require a memory-mapped tracking table. START-D increases cache misses by just 2.2\%  while START-LITE increases them by 2.6\% (including counter-misses). Across 51 single and mixed workloads, START-D's maximum slowdown is just 6.7\%, compared to 18.6\% for Hydra-544KB. Thus, START provides low-overhead protection for memory-intensive, cache-intensive, multi-programmed and multi-threaded workloads while avoiding significant dedicated storage structures.

\newpage
\bibliographystyle{IEEEtranS}
\bibliography{refs}

\end{document}